\documentclass[aps,prc,preprint,superscriptaddress,
amsfonts,amsmath,
amssymb,showpacs,floatfix]{revtex4}
\usepackage{graphicx,dcolumn,multirow}
\begin{document}
\title{High-${\boldsymbol j}$ single-particle neutron states outside the  
${\boldsymbol N}$=82 core}
\date{\today}
\author{B.P.~Kay}
\affiliation{Schuster Laboratory, University of Manchester, 
Manchester M13 9PL, UK}
\author{ \underline{S.J.~Freeman}}
\email[Correspondence to:~~]{sean.freeman@manchester.ac.uk}
\affiliation{Schuster Laboratory, University of Manchester, 
Manchester M13 9PL, UK}
\author{ J.P.~Schiffer}
\affiliation{Argonne National Laboratory,  Argonne,
Illinois 60439, USA}
\author{ J.A.~Clark}
\affiliation{A.W.Wright Nuclear Structure Laboratory, Yale University, 
New Haven, Connecticut 06520, USA}
\author{ C.~Deibel}
\affiliation{A.W.Wright Nuclear Structure Laboratory, Yale University, 
New Haven, Connecticut 06520, USA}
\author{ A.~Heinz}
\affiliation{A.W.Wright Nuclear Structure Laboratory, Yale University, 
New Haven, Connecticut 06520, USA}
\author{ A. Parikh}
\altaffiliation[Current address:~~]{Departament de Fisica i Enginyeria 
Nuclear, Universitat Politecnica de Catalunya,  
\hfil\break E-08036, Barcelona, Spain}
\affiliation{A.W.Wright Nuclear Structure Laboratory, Yale University, 
New Haven, Connecticut 06520, USA}
\author{ C.~Wrede}
\affiliation{A.W.Wright Nuclear Structure Laboratory, Yale University, 
New Haven, Connecticut 06520, USA}

\begin{abstract}
The behaviour of the $i_{13/2}$ and $h_{9/2}$ single-neutron strength 
was studied with the  ($\alpha$,$^{3}$He) reaction on $^{138}$Ba, 
$^{140}$Ce, $^{142}$Nd and
$^{144}$Sm targets at a beam energy of 51 MeV.
The separation between the single-neutron states $i_{13/2}$ and $h_{9/2}$ was
measured in $N$=83 nuclei with changing proton number.  To this end spectroscopic 
factors for states populated in high-$\ell$ transfer were extracted from 
the data.  Some mixing of $\ell$=5 and 6 strength was observed with states 
that are formed by coupling the $f_{7/2}$ state to the $2^+$ and $3^-$ vibrational 
states and the mixing matrix elements were found to be remarkably constant. 
The  centroids of the strength indicate a systematic  change in the 
energies of the $i_{13/2}$ and $h_{9/2}$ single-neutron states with increasing 
proton number that is in quantitative agreement with the effects expected from 
the tensor interaction.
\end{abstract}
\maketitle

The description of nuclear structure in terms of single nucleons moving 
within the nuclear mean field is at the root of 
much of the understanding of 
atomic nuclei. The properties of the available single-particle orbitals close 
to the Fermi surface are essential ingredients in the understanding of many 
phenomena, from the description of nuclear excitations within a shell model 
to the development of collectivity in low-lying vibrational states.
There is increasing experimental evidence, emerging 
 in situations where shell structure can be
tracked across a range of neutron excess,  
that the sequence 
of single-particle states is not  static and can change with neutron-to-proton 
ratio, indeed changing the `closed shells' in some regions of lighter nuclei
that are far from stability.  For example, changes in 
single-particle states occurring  in neutron-rich oxygen isotopes have been 
observed that destroy the familiar magic gap at $N=20$  close 
to the neutron dripline,
 whilst producing a new one at $N=16$ \cite{ozawa}. 
In heavier nuclei, systematic shifts in energies of
single-particle states have been observed, indicating similar
evolution of shell structure with changing neutron excess \cite{schiffer}.

Recently, Otsuka and co-workers identified the tensor part of the nucleon-nucleon
interaction as a general feature that appears to 
drive these observed trends in the evolution of nuclear shell structure in
different mass regions
\cite{otsuka2}. This is an interesting finding since, beyond the quadrupole 
moment of the deuteron, there have been
only a few instances where this inherent component of the nucleon-nucleon 
force manifests itself in nuclear structure in an obvious fashion \cite{fayache}.  
The tensor component of the nucleon-nucleon interaction has now been shown 
to have a systematic
effect on single-particle energies, an effect that can even 
cause the making and breaking of 
magic numbers \cite{otsuka2}. 
Efforts are now underway to include tensor components into
Hartree-Fock ({\sc hf}) mean-field models where they had been largely neglected 
\cite{otsuka3, brown, colo}.

Care needs to be taken when comparing theoretical predictions of single-particle 
energies with experimental data. 
Often little is known about the single-particle nature of the states in question. 
The energy of the lowest state of a given spin-parity can give misleading 
estimates for the energy of the
corresponding single-particle orbital if there is significant fragmentation of 
 strength.
Nucleon transfer reactions are a good probe of such fragmentation. However, 
even if experimental 
spectroscopic factors have been determined using such methods, their
absolute values are often of questionable meaning as they depend quite sensitively 
on choices made
in the calculation of reaction cross sections. 
For example, small changes in the bound-state radius of the transferred particle, or 
differences in optical-model potentials,
can have a dramatic effect on the magnitude of the calculated cross sections without 
appreciably changing the predicted angular distributions. 
The comparison of isolated experiments on different targets can therefore be fraught 
with difficulty arising from 
the details of the reaction models employed, in addition to any systematic 
experimental differences. 
Careful studies across all targets of interest, using the same experimental methods 
and the same choices in the extraction of spectroscopic factors, can alleviate 
most of these difficulties, at least in obtaining good relative spectroscopic factors. 
The kinematic characteristics of a particular transfer reaction also
influence the 
reliability of spectroscopic factors.
The assumptions inherent in the reaction models are best met where there is 
good momentum matching, ensuring relatively large cross sections for 
the direct transfer process.
Many studies have been performed on reactions, such as (d,p), which are well 
matched for low-$\ell$ transfer. 
However, there are fewer systematic studies where there is a large reaction 
Q value, and consequently
higher momentum transfer, leading to large cross sections for direct 
high-$\ell$ transfers. Such reactions 
 are therefore needed for reliable measurements of high-$j$ orbitals.

The work reported in Ref.~\cite{schiffer} describes the use of the ($\alpha$,t) 
reaction on stable tin targets to 
populate high-$j$ single-proton states in Sb isotopes. The relative spectroscopic 
factors for $\ell$=4 and 5
transfer were extracted using the same parameters in distorted-wave Born 
approximation  ({\sc dwba}) calculations 
and a common normalisation factor.
The results identifed the lowest 11/2$^{-}$ and 7/2$^{+}$ states with the dominant 
fragment of $h_{11/2}$ and $g_{7/2}$ 
single-proton states, with strengths constant to within $\pm10\%$. No single potential 
could describe the energetic separation 
of these states as a function of neutron number
until the introduction 
of a tensor term in the residual interaction \cite{otsuka2,otsuka3}.
 The variation of the lowest known experimental states with
the requisite spin and parity points to the possibility 
of a similar evolution of single-particle energies for 
neutrons outside $N$=82 \cite{schiffer}, 
in particular, for the $\nu i_{13/2}$ and $\nu h_{9/2}$ orbitals.
However, the single-particle nature
of the states involved is somewhat uncertain as no previous study of 
had been undertaken 
using reactions that are well matched for high-$\ell$ transfers.

Here we present a systematic study of spectroscopic factors across the stable metallic 
$N$=82 isotopes using the
 ($\alpha$,$^{3}$He) reaction, which has good kinematic matching for $\ell \sim 6$. 
 We have identified states populated 
 with high-$\ell$ transfer and extracted relative spectroscopic strengths 
  for $\ell$=5 and 6, allowing  
 the centroids of $i_{13/2}$ and $h_{9/2}$ strength to be determined.
  Since these states have no radial nodes, the comparison of their relative
energies is largely insensitive to anything other than specific features of the
residual interaction, as was the case for the states in Sb isotopes studied earlier.

A beam of $\alpha$ particles with an energy of 51 MeV was supplied by the Yale
tandem Van de Graaff accelerator
 and was used to bombard targets of $^{138}$Ba, $^{140}$Ce, 
$^{142}$Nd and $^{144}$Sm. 
Oxygen is an almost inevitable contaminant in such targets due to their
chemical reactivity.
To avoid complex handling procedures associated with self-supporting metallic 
targets, material was evaporated from enriched oxides onto 
supporting carbon foils of thickness 20--40~$\mu$g~cm$^{-2}$, which were then 
stored in an argon atmosphere with only brief exposure to air. Reactions on both 
oxygen and the carbon backing complicated the analysis, as discussed 
in detail below.
The Yale Enge split-pole spectrometer was used to momentum analyse the products 
from the reactions. A gas-filled ionisation chamber
and plastic scintillator, at the focal plane of the spectrometer, were 
used to isolate $^{3}$He ions from other
reaction products on the basis of their energies and energy-loss characteristics. A Si 
monitor detector was placed in the target 
chamber at 30$^\circ$ for diagnostic purposes. The beam dose was measured using a 
Faraday cup and beam-current 
integrator. The beam currents were typically in the range 10-30~pnA. Data were 
collected using the same experimental
set-up in two separate experimental runs.

In order to extract absolute cross sections, the product of the target thickness and 
spectrometer aperture was calibrated
using elastic scattering at low beam energies. In the first run this was done with a 
spectrometer angle of 30$^{\circ}$ at 
an incident energy of 10 MeV. In the second run, elastic scattering of 20-MeV 
$\alpha$ particles was measured at 20$^{\circ}$. 
The deviation from Rutherford scattering under both of these conditions
is predicted to be less than 0.5\% in 
 calculations using several different optical-model parameters.  
 The entrance aperture of the spectrometer was
 kept constant throughout each run to avoid any systematic differences. 
Using the nominal value of the entrance aperture, the $^{138}$Ba, $^{140}$Ce and 
$^{142}$Nd targets were found to be in the range
100-150~$\mu$g~cm$^{-2}$; the $^{144}$Sm target was thinner at 
$\sim$40~$\mu$g~cm$^{-2}$. 
No significant variation in the normalisation from yield to cross section was 
observed between measurements
made at the beginning and end of each run, and between the two runs, showing 
that the targets are fairly robust.

Typical momentum spectra for the observed ions are shown in Fig.~\ref{one},
where the main $N$=82 reaction strengths are via $\ell$=5 and 6 transfer, with
the exception of some $\ell$=3 transitions leading to the $7/2^-$ ground
state of the residual nucleus. An energy resolution of $\sim$70~keV was
obtained. Both $\ell$=5 and $\ell$=6 strengths are split into two groups in
each isotope. Also apparent in the spectra are intense lines arising from
reactions on $^{16}$O and $^{12}$C target contaminants. These contaminant
peaks obscure some of the lines of interest at certain angles. However, the
kinematic characteristics are such that the momenta of the ions
from contaminants change more quickly with angle than for the reactions on $N$=82
targets, allowing angles to be found where each peak of interest is free
from contamination. Weaker contaminant transitions from
$^{28}$Si($\alpha$,$^{3}$He) reactions were also present in some spectra, particularly
those from the $^{144}$Sm target. They were
easily identified using known excitation energies and from the momentum
variation with angle and were not close to any peaks of interest.

Data were taken at laboratory angles of 6$^{\circ}$, 11$^{\circ}$ and
20$^{\circ}$ for all targets. The first angle was chosen to be as close to
the peak yields for $\ell$=5 and 6 as possible without resulting in
count-rate problems, whilst 11$^{\circ}$ and 20$^{\circ}$ are angles
sensitive to the differences in the shapes of angular distributions. For the
$^{138}$Ba target, where contaminant ion groups cause most problems, spectra
were also taken at 30$^{\circ}$. Since spectroscopic factors are most reliable
when measured at the peak of the angular distribution at forward angles,
measurements were also made at 30$^{\circ}$ using the $^{140}$Ce
target as a consistency check on the reliability of spectroscopic factors
obtained at the larger angles.

Examples of angular distributions are shown in
Fig.~\ref{two}, together with {\sc dwba} calculations using optical-model
and bound-state parameters listed in Table~I.  The optical potentials were
successful in reproducing angular distributions of high-$\ell$ transitions
in Sn($\alpha$,t) reactions \cite{schiffer}. The exact finite-range
calculations were performed using the computer code {\sc ptolemy}
\cite{ptolemy}. The comparison with data indicates the quality of the {\sc
dwba} calculations in reproducing the angular variation in cross section.
Spin assignments were taken from the literature
\cite{NDS139,NDS141,NDS143,NDS145} although, in general, the measured
angular distributions agree with these spin assignments on the assumption
that $\ell$=5 and 6 transitions lead to the population of $9/2^{-}$ and
$13/2^{+}$ states, respectively. No strong transitions were observed with
transfers other than $\ell$=3, 5 and 6.

In $^{141}$Ce there is a doublet composed  of a 13/2$^{+}$ state at 1354.52(9)~keV 
and a 9/2$^{-}$ state at 1368.7(2)~keV \cite{NDS141} that are unresolved in the 
momentum spectra.  
Yields for these two states were extracted by fitting two Gaussian functions to the data
 with positions fixed using the known excitation energies. The widths were 
 fixed to values reproducing the shape of neighbouring singlet peaks, such that the 
 peak heights were the only free parameter in the fit. This procedure was successful 
 at 20$^\circ$ and 30$^\circ$. At 6$^\circ$ and 11$^\circ$  satisfactory fits were 
 not obtained due to an increasing background
in the spectra from inelastically scattered $\alpha$ particles that cannot be 
completely separated from ejectile $^{3}$He ions at forward angles. An 
alternative method, fitting two calculated $\ell$=5 and 6 angular distributions 
to the summed yield of the doublet led to ambiguous results, with a long flat 
trough in the $\chi^{2}$ space, due to the similarity in shape of the two 
distributions.

Table~II gives the differential cross sections measured in the current work, where
absolute numbers typically have an error of $\sim 7\%$ whilst relative numbers 
are good to $5 \%$.
Where a cross section value is not listed in the 
table, that particular angle was affected by contamination by oxygen or
carbon ion groups. In addition, data were not taken at 30$^{\circ}$ for
$^{142}$Nd and $^{144}$Sm targets and yields could not be extracted from the
$^{141}$Ce doublet at forward angles, as discussed above. 
Also shown are the spectroscopic
factors obtained using a single normalisation for both $\ell=5 {\rm~and~} 6$
transfers across all isotopes and transitions listed. The spectroscopic
factors at different angles typically varied by less than 10\% from the
average values listed in Table~II, apart from those arising from the Ce
doublet peak where the observed variation was less than the larger
errors arising from the fitting procedure. 
The relative values of the spectroscopic factors are subject to an uncertainty
of $\sim 10\%$, based on the error in the relative cross sections and
analyses performed using a variety of different sets of optical-model parameters.
Spectroscopic factors
were not extracted for the $\ell=3$ transitions to the residual ground
states; {\sc dwba} calculations were not able to reliably predict the shape
of these 
poorly-matched transfers. The spectroscopic factors show significant
deviations from those extracted previously using (d,p) reactions
(\cite{NDS139,NDS141,NDS143,NDS145} and references therein),
not only between different targets, but even for a given target with the
same $\ell$-value,
as might be
expected when comparing spectroscopic strengths extracted from
poorly-matched and well-matched transitions. 
The totals for $\ell$ =5 and for 6 strengths, shown in Table~III,
are constant to within $\pm 13 \%$ across the isotopes studied.

High-$\ell$ transfers populate two states for both $\ell=5$ and 6 in each final nucleus;
no other transitions were observed with these $\ell$ transfers
to a limit of 5\% of the strongest transition. 
This fragmentation is likely to be due to coupling to vibrational excitations of 
the core, leading to predominately two-state mixing between
 $0^{+}_{\rm core}\otimes \nu i_{13/2}$
 and $3^{-}_{\rm core}\otimes \nu f_{7/2}$ configurations in the case of the 
 positive-parity states, and $0^{+}_{\rm core}\otimes \nu h_{9/2}$ and 
 $2^{+}_{\rm core}\otimes \nu f_{7/2}$ configurations for the negative-parity
  states. Simple two-level mixing calculations have been used to extract the 
mixing matrix elements from the measured spectroscopic factors and excitation 
energies. The values of the matrix elements are shown in Table~III. They are
remarkably constant across all the nuclei studied, leaving the extent of the
mixing to be dictated by the energetic proximity of the unperturbed states,
 illustrated in
Fig.~\ref{three} (a) and (b). More detailed calculations 
with particle-core coupling models 
have been done in the past, where parameters 
were fitted to the available energies and spectroscopic 
factors \cite{heyde,trache,booth,oros}. 
The results of these calculations are broadly in line with the experimental 
results of the current work, although they have been fitted to spectroscopic 
factors from (d,p) reactions so a detailed comparison is not appropriate here.

The trend in the $i_{13/2}$ and $h_{9/2}$ 
centroids is similar to that 
of the lowest 9/2$^{-}$ and 13/2$^{+}$ states, as illustrated by the energy differences 
shown in Fig~\ref{three}(c). However, the fall in energy with increasing $Z$ is not 
as large and the crossing of the two lowest states is not observed in the 
single-particle energies.

In the $Z$=51 isotopes, the observed 
evolution of the 
$\pi h_{11/2}- \pi g_{7/2}$ gap as a function of neutron number, 
which is somewhat similar to the trends uncovered here, has been successfully 
reproduced by the introduction of the tensor-force terms in shell-model  and 
mean-field calculations \cite{otsuka2,otsuka3}. Qualitatively, these shifts have 
a fairly simple effect: nucleons  filling a $j=\ell\pm s$ orbit have an 
attractive effect with nucleons  in  $j=\ell \mp s$ orbitals, 
whereas a repulsive contribution to binding energy arises from the interaction 
between nucleons 
where both are either in  $j=\ell + s$ or $j=\ell - s$ states \cite{otsuka2}.

In the case of neutron states outside the $N$=82 core, the tensor force will
also produce such a shift
 arising from changes in the occupancy of various proton orbits.
Experimental proton occupancies in $N$=82 nuclei are available from previous 
systematic measurements of proton-transfer reactions \cite{wildenthal}; over the 
stable isotopes, the $\pi 1g_{7/2}$ and $\pi 2d_{5/2}$ orbits were observed to 
fill at approximately the same rate. These proton orbitals have opposite effects 
on neutron binding energies, having different spin-orbit couplings,
 $j=\ell -s$ and $j=\ell +s$. 
 But the effect of the $\pi 1g_{7/2}$ state should dominate, due to the 
considerably larger spatial overlap with the nodeless radial wavefunctions of the 
high-$j$ neutron states.
An increased binding energy for $i_{13/2}$ neutrons, and a decrease for 
$h_{9/2}$, should therefore result from  a tensor interaction with $g_{7/2}$ 
protons. As the occupancy of this proton orbital increases, the
two neutron states should drift towards each other in energy. 
This qualitative expectation is in agreement with the current experimental 
results, as illustrated in Fig.~\ref{three}.
More quantitatively, the effect of a tensor force based on $\pi$ and $\rho$ meson 
potential (discussed in detail in Ref.~\cite{otsuka2}) is to reduce the difference 
between $i_{13/2}$ and $h_{9/2}$ neutrons by 0.180~MeV per additional 
proton in the $\pi g_{7/2}$ orbital, using calculations of the relevant matrix elements
using $A$=140 harmonic oscillator wave functions \cite{otsuka_pc}.
  In line with the qualitative
expectations discussed above, the interaction with a $\pi d_{5/2}$ proton is 
calculated to
increase the gap by only 0.04~MeV. Combining these tensor matrix elements
with measured 
 proton occupancies \cite{wildenthal},
 the relative changes produced by 
the tensor interaction on the $\nu i_{13/2} - \nu h_{9/2}$ gap can be deduced. 
A comparison 
of this calculation with the experimental single-particle energy 
differences is shown in Fig.~\ref{three}(d), displaying reasonable agreement 
with the measured changes to the $\nu i_{13/2} - \nu h_{9/2}$ gap.

Calculations with mean-field models are more difficult in this case, compared to the
situation with proton orbitals outside $Z$=50, due to the 
need to reproduce the details of the filling of protons in the close-lying $\pi g_{7/2}$ 
and $\pi d_{5/2}$ orbitals near the Fermi surface. These states are known 
experimentally to lie only $\sim$100-300~keV apart \cite{wildenthal}. Recently 
{\sc hf} plus {\sc bcs}  calculations have been 
published for neutron states outside $N$=82  \cite{colo}, based on a modification
 to the  SLy5 Skyrme interaction that introduces elements from the tensor force. 
 It predicts a different pattern of filling proton orbitals compared to that deduced 
 experimentally, namely sequential filling of the $\pi g_{7/2}$ level, followed by 
 the $\pi d_{5/2}$ orbital. As a result, there are discrepancies between
 the calculated changes to the 
 $\nu i_{13/2} - \nu h_{9/2}$ gap as a function of $Z$ and 
 the experimental data presented here.

No quantitative information exists on the single-neutron nature of 
high-$j$ states in $N$=83 nuclei, other than those accessible in 
neutron-adding reactions 
on the stable metallic species. 
There are no $13/2^{+}$ states known in $^{137}$ Xe; 
measurements in normal kinematics for a $^{136}$Xe gas target are compromised 
by reactions on gas-cell windows leading to contamination problems.
 In lighter $N$=83 nuclides, direct experimental information 
about low-lying $13/2^{+}$ states is missing. However, the 
energies of both the {\it lowest} $13/2^{+}$ and 
$9/2^{-}$ states are known for $Z>62$, even though spectroscopic factors
are unavailable. Beyond $Z=62$ there is a 
sudden change in the energy 
systematics; the two states start to diverge in energy quite strongly 
(see Fig.~\ref{three}(c)). In these higher $Z$ systems,  the $\pi h_{11/2}$ state 
should begin to fill, with dramatic consequences for high-$j$ neutron states due 
to their large radial overlap and an increasing repulsive contribution from a 
tensor interaction, amounting to 0.16~MeV per additional proton \cite{otsuka_pc}. 
However, any strong conclusions should be 
 treated with some caution since the single-particle content of 
the lowest-lying states will not be known until nucleon-adding reactions are 
performed on radioactive beams. Indeed, it is towards the $Z$=64 sub-shell 
closure that the $3^{-}$ core excitation is expected to fall closest in energy to 
the centroid of $i_{13/2}$ strength (see Fig.~\ref{three}(b)). Assuming 
the mixing matrix elements are of similar strength to those extracted here, 
this will lead to large 
mixing and significant fragmentation of $\ell$=6 strength between two $13/2^{+}$
states.  The energy of the lowest $13/2^+$ state will therefore be a particularly
unreliable
estimate of the $i_{13/2}$ 
single-neutron energy.

In conclusion, spectroscopic factors have been measured for high-$j$ states outside 
stable $N$=82 isotopes revealing a systematic drift in $i_{13/2}$ and 
$h_{9/2}$ single-neutron energies toward each 
other with increasing proton number. There is quantitative agreement between the 
experimental trends and the shifts in 
single-particle energies predicted with a 
tensor interaction.  

We are indebted to Takaharu
Otsuka for providing us with the relevant tensor matrix elements 
 and for many very fruitful discussions.
We would like to acknowledge meticulous
target manufacture by John Greene (Argonne National Laboratory)
and Paul Morrall (Daresbury Laboratory). 
Peter Parker has provided invaluable assistance and advice,
without which these measurements would not have been 
possible. We would also like to 
thank the staff at Yale University, in particular, their
 efficient facility operation.
 This work was supported by the UK
 Science and Technology Facilities Council and the  
US Department of Energy, Office of Nuclear Physics,
 under Contract Numbers DE-FG02-91ER-40609 and DE-AC02-06CH11357.

\vfil\eject

\begin{figure}
\caption{\label{one} Spectra from the ($\alpha$,$^{3}$He) reaction 
as a function of excitation energy taken at 20$^{\circ}$ in the laboratory frame, labelled 
by residual nucleus. Peaks are labelled with the $\ell$ transfer of the transition. 
The large peak moving up through the spectra with increasing atomic number is 
due to reactions  carbon and oxygen contaminants in the target. Small, broader peaks, 
particularly evident in the $^{145}$Sm spectrum, arise from silicon contamination.}
\end{figure}
\begin{figure}
\caption{\label{two} 
Differential cross sections from the
$^{144}$Sm($\alpha$,$^{3}$He)$^{145}$Sm reaction shown for transitions
leading
to the following states in the residual system: (a) 13/2$^{+}$ state at
1105~keV,
(b) 9/2$^{-}$ state at 1423~keV, (c) 13/2$^{+}$ state at 2670~keV and
(d)  9/2$^{-}$ state at 1780~keV. Square points indicate $\ell=5$
transitions, circular points $\ell=6$ assignments. The lines are {\sc
dwba} calculations normalised to the data, for $\ell$=5 (continuous line)
and for $\ell$=6 (dashed line).
 }
\end{figure}
\begin{figure}
\caption{\label{three} (a) Experimental centroid energies of the neutron 
$h_{9/2}$ single-particle strength for $N=83$ nuclei (solid squares)
compared to the quadrupole vibrational excitation energies of the corresponding 
even-even core (open symbols). (b) A similar plot for the $i_{13/2}$ 
centroid and octupole vibrations of the core. 
The difference in experimental centroid 
energies of  $\nu i_{13/2}$ and $\nu h_{9/2}$ strength (solid triangles) compared 
to: (c) the energy differences between the lowest 13/2$^{+}$ and 9/2$^{-}$ states 
(open triangles),  and (d) predictions of a $\pi+\rho$ tensor interaction
(grey band). The width of the 
grey band is generated by experimental uncertainty in the proton 
occupancies \cite{wildenthal}.  }
\end{figure}  

\begingroup
\begin{table}
           \label{tone}
   \caption{ Optical-model  and bound-state potentials used in the {\sc dwba} 
   calculations. Depths of the bound-state potentials, marked with $\star$, were 
   varied to reproduce the appropriate binding energies.\\
   }
   \centering
    \begin{tabular}{lccccccccccr} 
    \hline
     \hline
  \multirow{2}{*}{Channel~~~~~~} &~~~$V$~~~&~~~$r_{r}$~~~&~~~$a_{r}$~~~ & ~~~
  $W$~~~ &~~~ $r_{i}$ ~~~& ~~~$a_{i}$~~~&~~~$V_{so}$ ~~~&~~~ 
  $r_{so}$ ~~~&~~~ $a_{so}$~~~& \multirow{2}{*}{~~~Reference} \\
               & MeV    & fm  & fm  & MeV & fm & fm & MeV &fm& fm &  \\
      \hline\hline
      $^{4}$He & 207.0&1.3&0.65&28.0&1.3&0.52&--&--&--&\cite{bassani}\\
      $^{3}$He & 152.0& 1.24& 0.685& 23.0&1.432&0.870&--&--&--&\cite{flynn}\\
      $N$=82+n &$\star$&1.25&0.63&--&--&--&7.0&1.10&0.50&\cite{parkinson}\\
      $^{3}$He+n&$\star$&1.20&0.65&--&--&--&--&--&--&\cite{low}\\
   \hline\hline

  \end{tabular}   

\end{table}
\begin{table}
           \label{ttwo}
   \caption{Differential cross sections and spectroscopic factors of states 
   populated by large $\ell$ transfer in the ($\alpha$,$^{3}$He) reaction 
   on  stable $N$=82 targets.  Excitation energies are taken from 
   Ref.~\cite{NDS139,NDS141,NDS143,NDS145}.\\}

   \centering
 \begin{tabular}{||c||cc|cccc|c||} 
      \hline\hline
   & Excitation &\multirow{2}{*}{~~~~$\ell$~~~~} &\multicolumn{3}{c}{Cross Section (mb/sr)}&&\multirow{2}{*}{~~~~
   $C^{2}S$~~~~} \\
        &Energy (keV)&&~~~~~~6$^{\circ}$~~~~&~~~~
        11$^{\circ}$~~~~&~~~~20$^{\circ}$~~~~&~~~~30$^{\circ}$~~~~&~~~~\\
    \hline\hline
    $^{139}$Ba	&1283.32(3)	&5	&--		&--		&0.35	&0.20	&0.70\\
    			&1538.96(11)	&6	&--		&--		&0.75	&0.44	&0.60\\
    			&1619(10)	&5	&--		&--		&0.18	&0.11	&0.41\\
    			&3080(10)	&6	&0.27	&0.21	&0.12 	&0.08	&0.17\\
\hline
     $^{141}$Ce&1354.52(9)	&5&--		&--		&0.54	&0.21	&0.67\\
    			&1368.7(2)	&6&--		&--		&1.01	&0.57	&0.79\\
    			&1693.3(1)	&5&--		&0.15	&0.13	&0.08	&0.25\\
    			&2899(2)		&6&0.44		&0.31	&--		&0.15	&0.22\\
    \hline

       $^{143}$Nd&1228.04(8)	&6	&2.68	&2.06	&1.19	&--&0.65\\
    			&1407.08(6)	&5	&1.11	&0.76	&0.54	&--&0.83\\
   			&1739.21(8)	&5	&0.34	&0.22	&0.16	&--&0.29\\
    			&2805.3(3)	&6	&--		&--		&0.26	&--&0.22\\
    
    \hline
    $^{145}$Sm	&1105.3(16)	&6	&2.99	&2.30	&1.25	&--&0.66\\
    			&1423.24(3)	&5	&1.18	&0.87		&0.62	&--&0.84\\
    			&1780.32(9)	&5	&0.41	  &0.29		&0.22	&--&0.34\\
    			&2670.0(11)	&6	&0.87 	&0.62		&0.41	&--&0.30\\
     \hline\hline
  \end{tabular}

\end{table}

\begin{table}
           \label{tthree}
   \caption{Summed single-neutron strengths, energy centroids and mixing
matrix elements with the state of corresponding $j^{\pi}$ formed by
coupling the $\nu f_{7/2}$ state to the $2^+$ or $3^-$ core vibrations.\\}

  \resizebox{!}{2.25cm} 
    {\begin{tabular}{lcccccccc} 
    \hline
     \hline
    &\multicolumn{2}{c}{$^{139}$Ba}&\multicolumn{2}{c}{$^{141}$Ce}&
    \multicolumn{2}{c}{$^{143}$Nd}&\multicolumn{2}{c}{$^{145}$Sm}\\
     &~~~~h$_{9/2}$~~~~&~~~~i$_{13/2}$~~~~&~~~~h$_{9/2}$	~~~~&~~~~
     i$_{13/2}$~~~~	&~~~~h$_{9/2}$~~~~	&~~~~i$_{13/2}$~~~~	&~~~~
     h$_{9/2}$~~~~	&~~~~i$_{13/2}$~~~~	\\
     \hline
   $\sum C^{2}S$& 1.11(16)&0.77(11)&0.92(13)&1.01(14)&1.12(16)
   &0.87(12)&1.18(17)&0.96(14)\\
    Centroid Energy (keV)&1407(10)&1879(24)&1447(10)&1702(52)
    &1493(5)&1627(31)&1526(10)&1594(29)\\
  \hline
   \%  Upper State&37(2)&22(1)&27(3)&22(3)&26(1)&25(2)&29(3)&31(1)\\
  Matrix Element (keV)&162(6)&639(17)&151(5)&632(34)&145(2)
  &686(16)&162(4)&725(9)\\
        \hline \hline
     \end{tabular} }  
  \end{table}

\endgroup

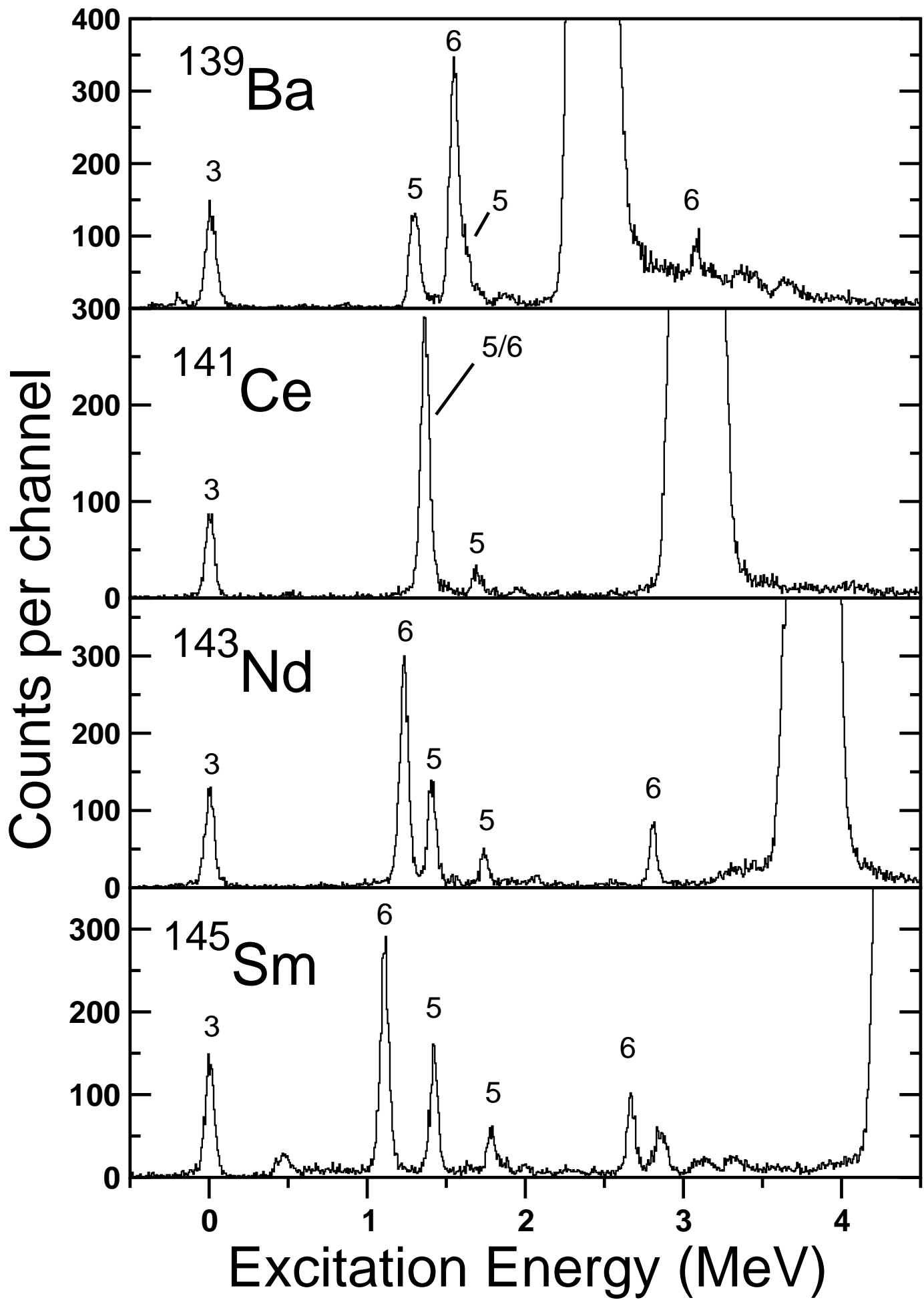
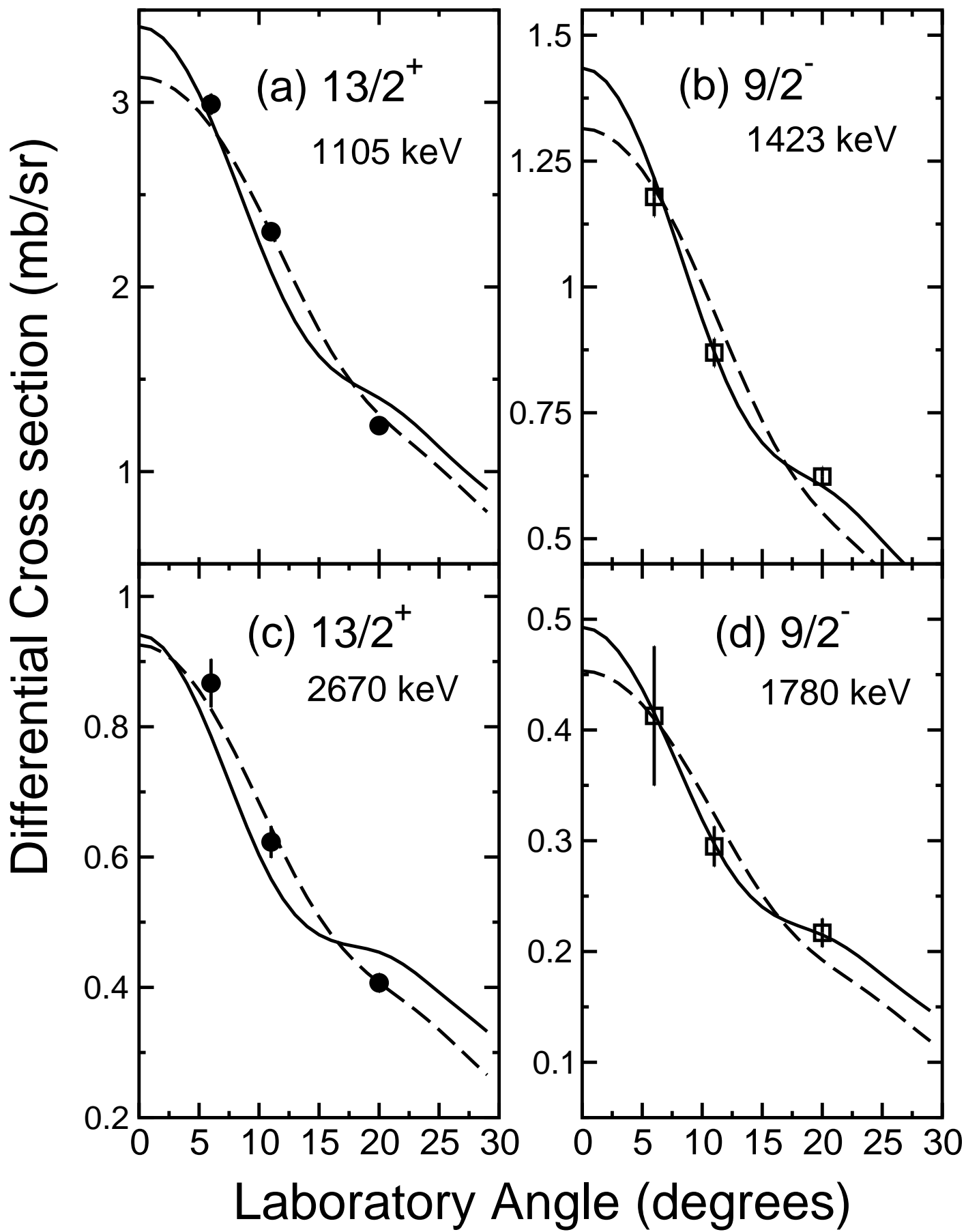
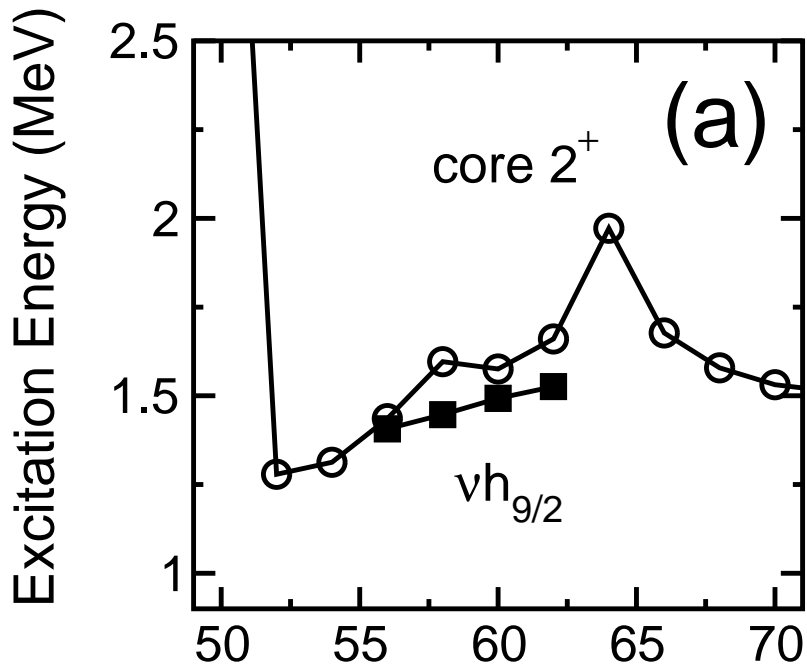
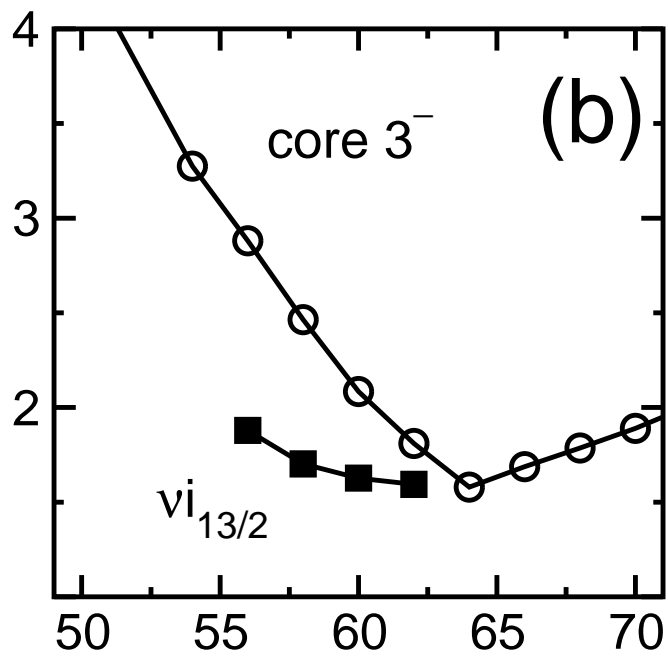
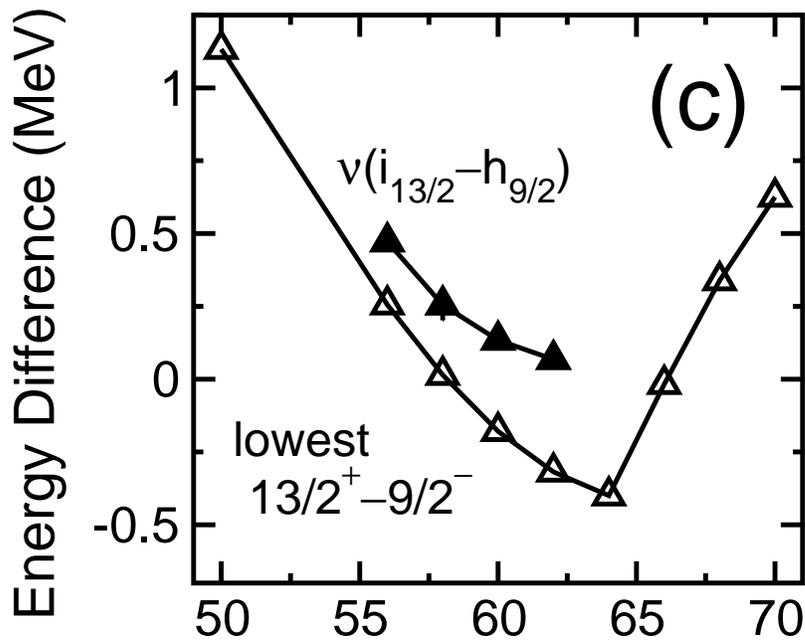
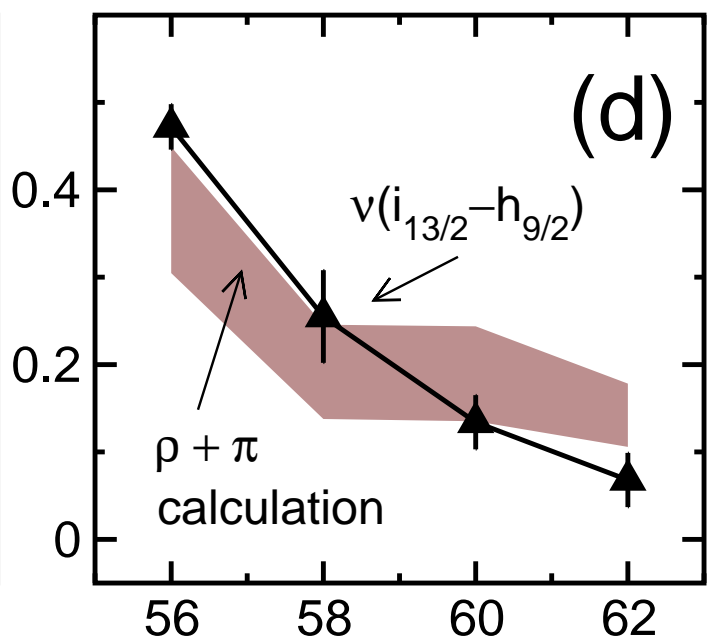
\end{document}